\pdfoutput=1 
\documentclass[10pt,a4paper]{article}
\usepackage[utf8]{inputenc}
\usepackage[T1]{fontenc}
\usepackage[english]{babel}
\usepackage{amsmath}
\usepackage{amsfonts}
\usepackage{amssymb}
\usepackage{graphicx}

\usepackage[parfill]{parskip} 
\usepackage{csquotes}
\usepackage{hyperref}
\hypersetup{linktoc=page}
\usepackage{float}
\usepackage{subcaption}
\usepackage[most]{tcolorbox}
\usepackage{booktabs}
\usepackage{enumitem}
\usepackage{tocloft}

\usepackage[backend=biber,style=alphabetic]{biblatex}
\bibliography{timestampsavings}

\author{Andrea Barontini\thanks{andrea.barontini@bybaro.it}}
\title{Notarial timestamps savings in logs management via Merkle trees and Key Derivation Functions}
\date{September, 2021}

\begin{document}
	\maketitle
	
	\begin{abstract}
		Nowadays log files handling imposes to ISPs (intended in their widest scope) strict normative duties apart from common technological issues. This work analyses how retention time policies and timestamping are deeply interlinked from the point of view of service providers, possibly leading to costs rise. A new schema is proposed trying to mitigate the need for third-party suppliers, enforcing cryptographic primitives well established in other fields of Information Technology but perhaps not yet widespread in logs management. The foundations of timestamping are recapped, and properties of cryptographic primitives introduced as a natural way to bypass legacy schema inefficiency and as an extra level of protection: these choices are justified by savings estimation (with regard to different ISP magnitudes) and by some basic security considerations.
	\end{abstract}

	\vskip 1cm

	\section{The context}
	In the last years consciousness and concerns about privacy have raised, data handling by Internet and telecommunication providers is under the spotlight as never before, lawmakers are eventually entering in the field, a fair balance between cloud services convenience and worries about digital identities is far from being reached.
	
	It's easy to understand that applications, servers, network logs are no more just an IT operations' commodity useful to troubleshoot technical problems, but duties about their management are emerging; the landscape is various, and this work deals only with a couple of very basic requirements and their link:
	\begin{itemize}
		\item \textit{retention times}, i.e. the minimum amount of time a log must be available upon request and the maximum amount of time it can be conserved (e.g.: police can ask a provider for some data one week after it has been recorded, but after -let's say- five years that data must be deleted);
		\item \textit{tamper-proof-ness}, meaning that data since recorded cannot be changed without invalidating it.
	\end{itemize}
    Retention times are usually enforced by disincentives to circumvent their requirements: if service owner is discovered without needed data or taking advantage of data later than its retention time limit, it should be sanctioned. To be able to claim for data integrity, a technological proof in the form of a \textit{notarized timestamp} of the log file is instead required, given how easily an unfaithful service provider could alter the logs forging a new file (how it's really easier than deleting a log only far after the time limit is a moot point, but that's how the world goes). It's common to apply timestamps during log files rotation.
    
    It's worth noting that retention times actually are a property of each event recorded in a log file,  not of the log file as a whole: for example on a network device a root login and an interface UP/DOWN will probably have very different retention times requirements even if they could be collected in the same file. It means that, even considering a data-lake aggregation strategy producing a single huge log file (and that's not always the case), a per-event-type disaggregation is anyway needed to separately timestamp different files, each containing only events sharing the same retention policies: this way all conservation time constraints can be respected without breaking any timestamp. 
    
    Organizations under a legislation (or with a security department) mandating many different retentions, or not being able to process their logs as a unique data-lake, or needing frequent rotations due to data amount could easily see a growing number of needed notarized timestamps: they often are a paid third-party service, so it makes sense to propose a way to reduce that quantity.
    
    \section{Timestamps fundamentals}
    Notarization of a file  is not only a technological matter: dealing with legal validity it's of course also driven by law constraints: for example lawmakers could value guarantees coming from externalities, e.g. emitting organization's assets. This kind of choices influences the type of infrastructure and (de)centralization level of the issuer \cite{tskinds}; some examples, without pretending completeness:
    \begin{itemize}
    	\item timestamps based on a Proof-of-Work blockchain with granular-enough blocks mining schedule (please note that even if fully decentralized this case is not free of charge because of transactions fees) \cite{blocknot};
    	\item timestamps as digital certificates delivered to the  applier by an accredited issuer \cite{RFC3161};
    	\item timestamps as an entry in a carefully protected database managed by an organization guaranteeing its existence for a certain number of years \cite{bucap};
    	\item \dots
    	\item some mix of the previous ones.
    \end{itemize}
    However all solutions have a common technological ground: to save space all of them use a digest of the file produced by a \textit{cryptographic hash} (we will simply refer to it as \enquote{hash output} or even as just \enquote{hash}) and associate it to a time tag (writing the digest in a block at a specific height referable to that time, signing the association, storing the association in a secure DB, \dots).
    
    A cryptographic hash is employed because \cite{preimage}:
     \begin{itemize}
    	\item\textit{preimage resistance} guarantees that a file to be notarized must exist before timestamp emission, because we cannot timestamp a fake hash output, and only later forge a document corresponding to the fake hash (put it simply, the hash is \textit{one-way} so we cannot invert it);
    	\item \textit{second-preimage resistance} makes impossible, since a document is hashed and timestamped, to forge a second different document with the same hash (so we cannot pretend to have timestamped a document different from the original one).
    \end{itemize}
    As background for next sections considerations, it's important to note that:
    \begin{itemize}
		\item the foregoing hash properties are \textit{computational properties}, meaning they hold because of limits of our computation resources: having an unrealistically fast computer  or -equally- an unrealistic long time in hand we could break them by brute force (trying any hash input until success).
		\item from a general point of view we cannot say if conditioning hash input to be a valid log file (so considering only \enquote{well-formed} inputs, respecting log grammar/syntax) strengthen or weaken the hash properties: roughly speaking it depends on how hash output  distribution \enquote{reacts} to the restricted case. Regarding brute force attacks, the input restriction for sure could reduce the domain; however tries-set would still be so \enquote{big} to get searched that it's not obviously advantageous to also add the overhead of a priori input selection, considering how fast hashing is. Anyway we will be only interested in comparing a new timestamping schema to the current one: so we will consider their relative security, ignoring their effectiveness with respect to general hash theory and assuming security of the current timestamping procedure as an accepted matter of fact.
	\end{itemize}
	In the light of what we have said, we now know that to separately  timestamp many files we usually invoke the timestamping service many independent times, each time for a different hash: so to gain savings we need to reduce the number of third-party-processed hashes, trying at the same time to not lose mutual independence of each file's \textit{timestamp markers} (please note that now we have added the \enquote{markers} qualifier to underline a more complex structure: if the number of processed hashes will be lower than the number of files, emitted timestamps will not be enough -by themselves- for each file, so we'll also need something else).
	
	\section{Merkle trees as  Accumulators}
	The \enquote{to reduce the number of third-party-processed hashes} above is actually an euphemism: it's possible to use just one hash to timestamp $n$ files (which of course will share the time certification, nightly rotated logs are good candidates). To get this result we exploit the accumulator nature of Merkle trees \cite{merkletree}:
	\begin{itemize}
		\item in Merkle trees each internal node is the hash of its two children nodes \textit{concatenated}; each leaf node is instead the hash of one of the tree's inputs (in our case, the log files). If an internal node has only one child (situation occurring when cardinality of input-set is not a power of 2) the single child is concatenated with itself. The top level node is called \textit{tree root}. E.g.:
		\begin{figure}[h!]
			\includegraphics[width=\linewidth]{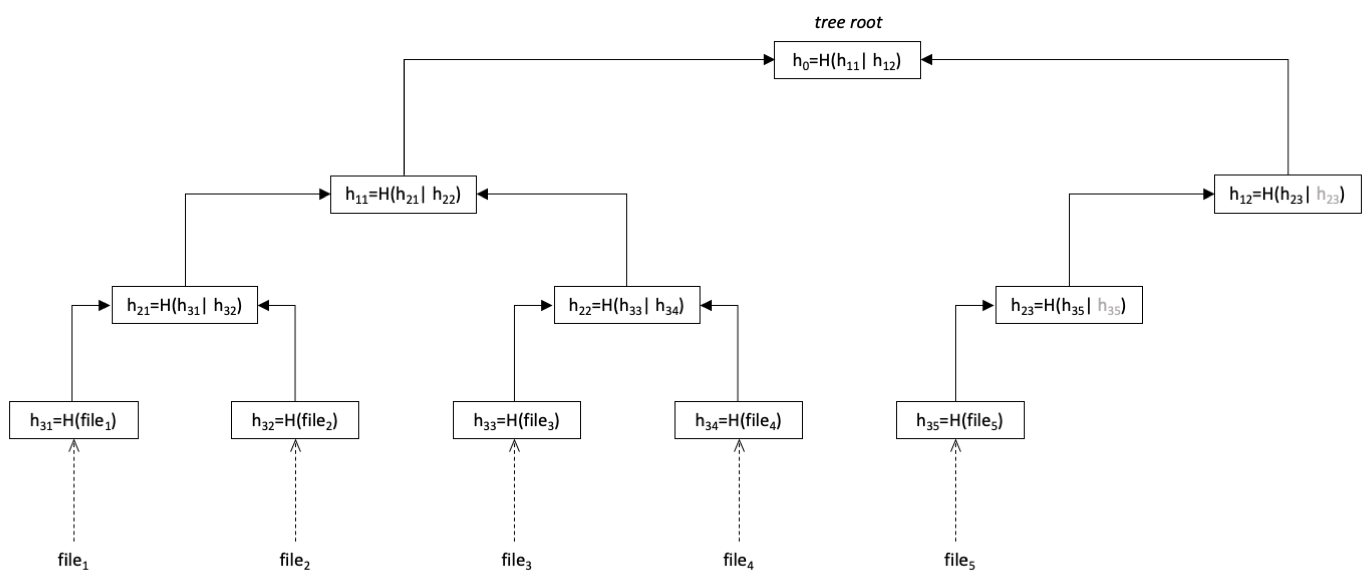}
		\end{figure}
	
		Of course with 6 input files we would have $h_{23}=H(h_{35} | h_{36})$ and $h_{12}$ would be the only node with just one single child.
		\item It's easy to get convinced that \enquote{membership} of an input file in a Merkle tree can always be proved knowing the tree root representing the whole tree and $\left\lceil \log_{2}n \right\rceil$ other nodes values $h_{yx}$ (where $n$ is the number of inputs): that's the accumulator nature of Merkle trees. E.g.: to prove membership of \textit{file}$_3$ in our example tree we need to use file contents and the black nodes values to calculate the tree root and check if it is equal to the one we pre-know:
		\begin{figure}[h!]
			\includegraphics[width=\linewidth]{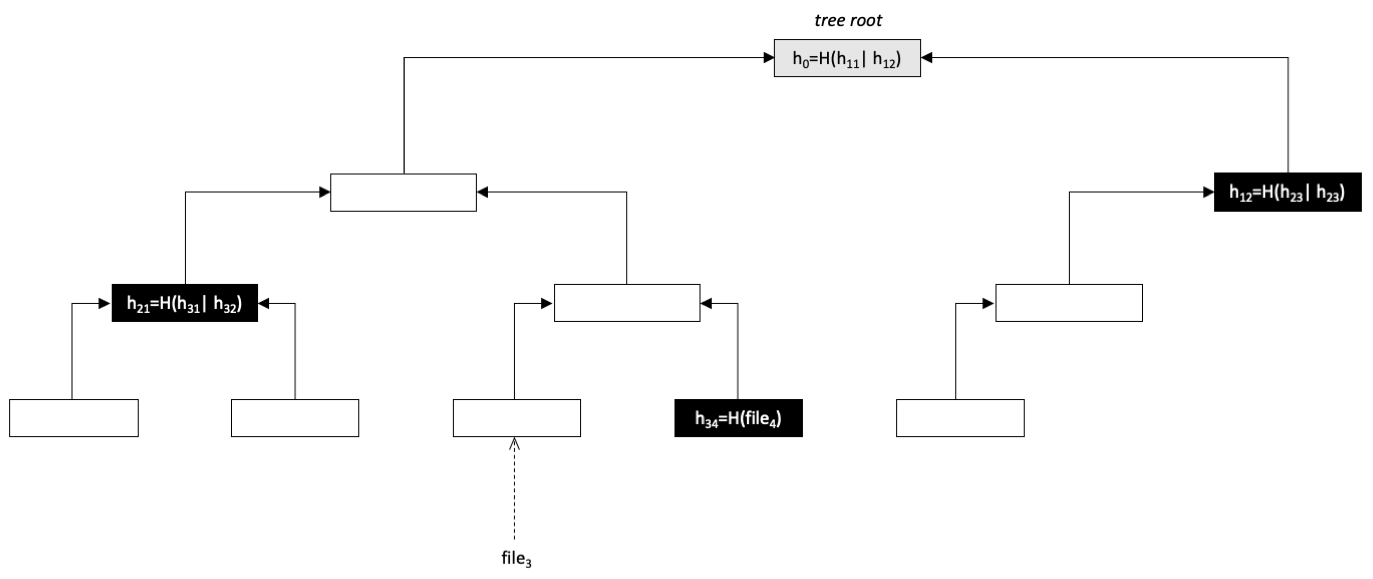}
		\end{figure}
	
		The ordered set (from inputs side to tree root) of black nodes values is called \textit{Merkle path}.
		
		Lastly, please note that check of inputs whose way towards tree root meets single-child nodes (e.g. our \textit{file}$_5$) could need less values; however to avoid dealing with special cases we choose to use the \enquote{canonical} proof style for them as well (repeating the same value when needed).
	\end{itemize}

	\section{The new timestamping schema and its savings}
    We now have all the elements to go back to our original problem: how to save timestamps. At this point the strategy should be self-evident:
    \begin{itemize}
    	\item the only timestamped hash will be the tree root;
    	\item for each log file, the earlier minted \textit{timestamp markers} will be made up of the same, unique timestamped tree root and file's Merkle path;
    	\item to prove a file hasn't been changed since timestamping, we check that Merkle root calculated from its content and Merkle path is  equal to the previously timestamped one.
    \end{itemize}
	It seems a win-win situation: let's compare legacy and new schema
	(we assume one logs rotation per-day, 32 bytes SHA256 hashes, Timestamp Authority TSA providing timestamps as X.509 certificates with common size of about 5 KBytes; per-timestamp price comes from a quick search of bundles available on the market during Q3 2021 \cite{ufficiocame}; needed storage values are rounded to the greatest-less-than-themselves Byte's $2^{10}$ multiples to be more readable; furthermore they take into account just timestamps or timestamp markers occupations, of course not source log files):
	
	\begin{figure}[h!]
		\includegraphics[width=\linewidth]{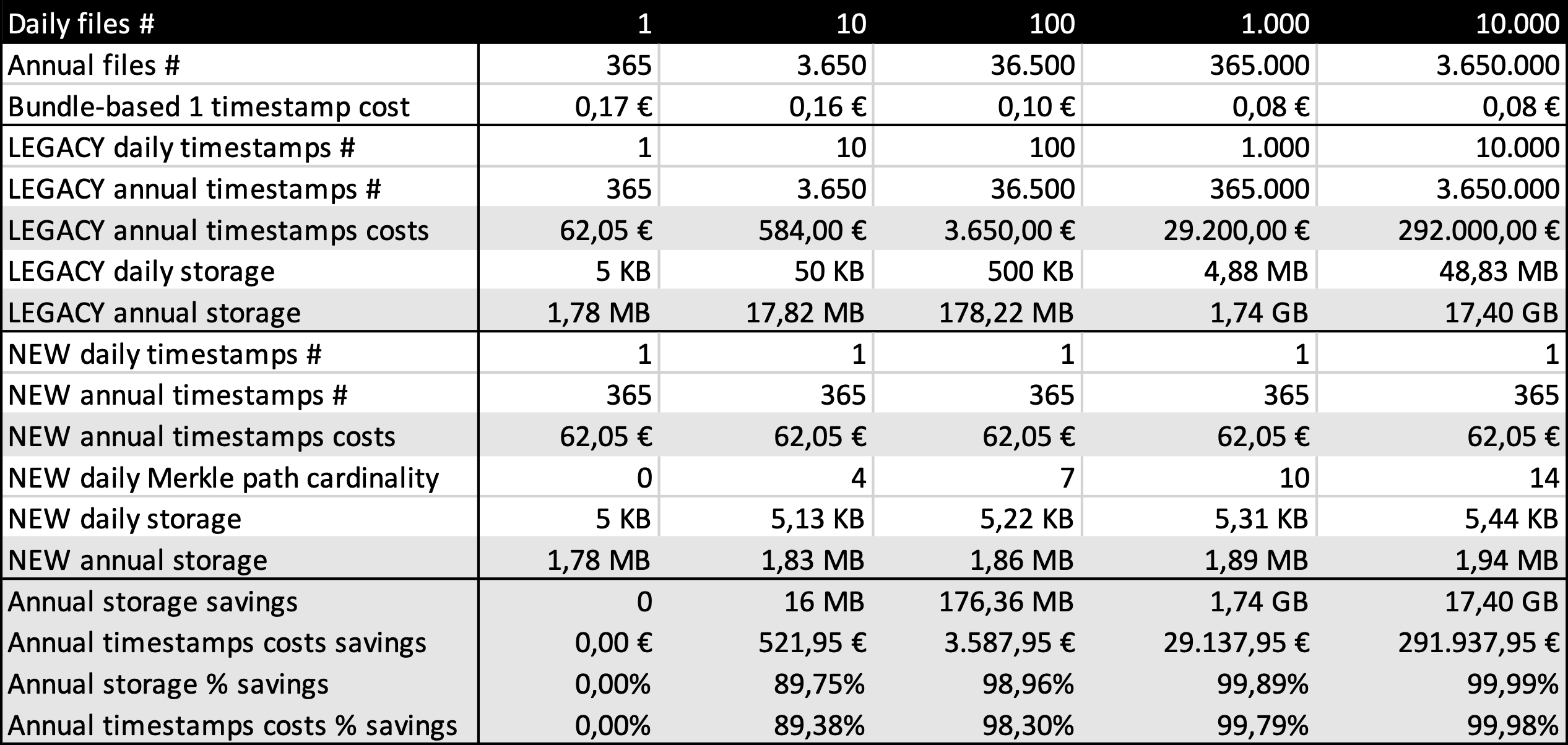}
	\end{figure}

	\section{Security considerations}
	An easy (and someway correct) objection to the proposed schema is that Merkle paths provide search space for a cheating log-manager trying to post-forge a fake file after timestamping procedure. Calculation of Merkle root during verification can be seen as:
	\begin{equation*}
		\text{Merkle root} = H( f(\text{Merkle path}, \text{log file}) )
	\end{equation*}
	No matter if $f$ function contains hashes $H$, as well: now from our point of view it's just a function, with log file and Merkle path as inputs. Note that the schema doesn't require the Merkle path to be advertised earlier than timestamp verification phase, so a cheating log-manager could try a \textit{second-preimage attack} to the hash function:
	\begin{equation*}
		H( f(\text{Merkle path}, \text{log file}) ) = H( f(\text{fake Merkle path}, \text{forged log file}) ) 
	\end{equation*}
	The fake Merkle path will offer a relevant search space: considering that \textit{each} of its elements is itself an hash output, the input bits to play with -once a forged log file has been chosen- are a multiple of outer $H$ codomain size.
	
	It's known \textit{collision resistance} is weaker than \textit{second-preimage} and \textit{preimage resistance}, given that in the former we can choose both preimages: so if we discover it's not affordable to find a collision, then a stronger difficulty limits the other attacks. Literature (i.e. Birthday paradox \cite{birthdayp} \cite{birthdaya}) states that the expected number $Q$ of tries to find an hash collision is proportional to the square root of the hash codomain. Considering SHA256:
	\begin{equation*}
		Q \approx \sqrt{2^{256}} = 2^{128}
	\end{equation*}
	Let's try to understand what $2^{128}$ tries means with an example.
	
	The maximum time a cheating log-manager could have to find a collision is given by the sum of the log retention time and of the access-request maximum handling time (in the unlikely case the search begins soon after the log collection, the access request is made near the end of the retention time, and all permitted time is used to deal with the request). 
	Italian Telephone and Internet Service Providers can reach 6 years retention of metadata; it's common to store so old logs offline, so let's consider 30 days for their handling: more or less $2^{28}$ seconds overall. Succeeding to cheat within this time means trying  $2^{128}/2^{28}=2^{100}$ fake Merkle paths per-second. Is it possible? 
	
	State-of-the-art in hash computation (double round hash actually) is currently -Q3 2021- represented by Bitcoin ASIC miners, delivering 100 TH/s \cite{bitmain}: about $2^{47}$ double round hashes per-second. Even \enquote{forgetting} that calculation of Merkle root involves multiple hashes, it's evident that today computational power doesn't allow collisions under the given assumptions (so, as said earlier, we are safe from second-preimage attacks as well). 
	
	With the foregoing result it's also worth noting that we don't strictly need a reduction from \textit{new timestamp markers} attacks to \textit{legacy timestamps} attacks (or, seeing it from a more technical point of view, from \textit{Merkle root} attacks to \textit{hash} attacks), because we have seen it's unfeasible to exploit the new extra possibilities of attack, if any (and it's not a surprise because $Q$ depends only on hash codomain size, so it's the same in the two cases).
	
	The \enquote{given assumptions} statement can be a risky point when dealing with security, so let's explicitly state ours: the hashes are modeled as \textit{Random Oracles} \cite{rom} returning \textit{random} and \textit{uniform} outputs. Unluckily actual employed hashes are not ideal \textit{ROs}, anyway a pragmatic approach about that is widely accepted (at least in Information Technology field where, e.g., SHA256 are not under discussion and Merkle tree structures have an important role in Bitcoin decentralized consensus process \cite{masteringmerkle}).
	
	\section{Schema improvement}
    Given the foregoing security considerations, we could take some extra precautions to complicate (and so discourage) brute force attacks:
    \begin{itemize}
		\item we could make cheaters' life more difficult by using huge codomain hashes, but computational capabilities are always improving and needing to periodically change hash type to preserve security level wouldn't seem a feasible time-proof approach;
		\item better to use a tuneable solution using a slowing Key Derivation Function \cite{kdf} \enquote{between} Merkle root and third-party timestamping service: imposing enough KDF repetitions, it would take longer to try enough fake Merkle paths during an attack:
	\end{itemize}

	\begin{figure}[h!]
		\includegraphics[width=\linewidth]{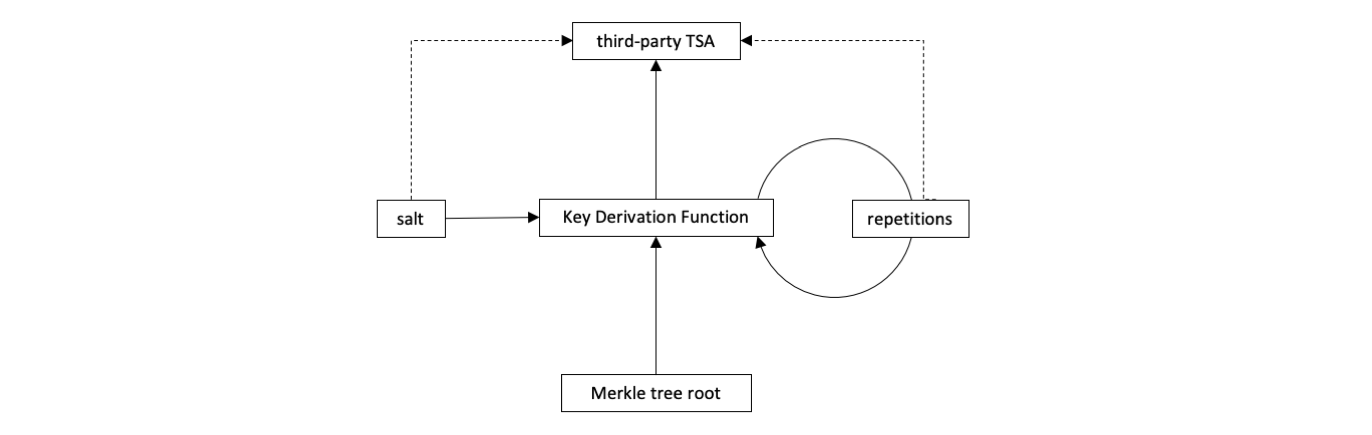}
	\end{figure}

	The \textit{salt} is a random value avoiding attacks by means of pre-computed rainbow tables, while \textit{repetitions} is the number of KDF iterations (so the slowness tuneable parameter):
	\begin{itemize}
		\item  both of them (or any other KDF parameter) could also be committed, by means of an extra hash -not explicitly shown in the picture- compressing them and KDF output before TSA processing, so to constrain their values;
		\item their retention needs only minor increase in storage requirements.
	\end{itemize}
	It's not between the aims of this work to give an indication about which Key Derivation Function to use (and with which parameters), anyway it seems right to cite an often adopted one, \textit{PBKDF2} (used in BIP39 Bitcoin wallets, for example \cite{masteringpbkdf2}), and the \enquote{new} (2015) kid-in-town \textit{Argon2}. Their parameters should be chosen to enforce the maximum process delay compatible with logs timestamping allowed delay (e.g. if logs have to be timestamped in 10 minutes since rotation, KDF parameters cannot be set to force a 15 minutes delay on process); and to find this compromise in term of KDF parameters, conservative assumptions have to be made about computational resources of log collectors at the time of timestamping (this conservative estimation together with usual computational capabilities improvement over time could result in a decreasing effectiveness of process delay, anyway an overall speed reduction would be attained).
	
	Lastly, the number $n$ of timestamped files can also be committed in the same way of \textit{salt} and  \textit{repetitions}: remembering it determines the cardinality of Merkle path, it imposes a constraint on the size of its representation as bit-string, so reducing attacker's degrees of freedom.
	
	\printbibliography
\end{document}